\documentclass[seceq]{ptptex}
\usepackage{graphicx} 
\usepackage{latexsym}
\notypesetlogo
\begin{document}
  \begin{flushright}
   RUP-05-2 \\
   September, 2005
\end{flushright}
\vspace{10mm} 
\begin{center}
\Large{\bf  Factorization Algorithm for Parton Showers \\ beyond the Leading Logarithmic Order of QCD  }

\vspace{15mm}

\large{Hidekazu {\sc Tanaka},$^1$ Tetsuya {\sc Sugiura}$^2$ and Yuya {\sc Wakabayashi}$^1$ \\

$^1$Department of Physics, Rikkyo University, Tokyo 171-8501, Japan \\
$^2$Waseda High School, Tokyo 162-8654, Japan
     }
     
  \vspace{25mm}

      {\Large ABSTRACT}
  \end{center}
  
  \vspace{10mm}
  
 A factorization algorithm for a patron shower model based on the evolution of momentum distributions proposed in a previous work is studied.
The scaling violation of initial state parton distributions is generated using parton showers to an accuracy of the next-to-leading logarithmic (NLL) order of quantum chromodynamics (QCD) using the information from only splitting functions and initial parton distributions at some fixed low energy. 
 In the algorithm proposed in this paper, the total momentum of the initial state partons is conserved in any factorization scheme for mass singularities. 
 As an example, the scaling violation of the parton distributions and the transverse momentum distributions due to initial state parton radiation are calculated with the jet calculs scheme. 

\section{Introduction}

 In perturbative quantum chromodynamics (QCD), logarithmic contributions due to collinear parton production are subtracted from a hard scattering cross section, and these contributions are absorbed into the parton distributions of a hadron.   
 In higher order calculations, the factorized terms from the hard scattering cross section are compensated for by associated terms of the splitting functions.  Therefore, the scaling violation of the parton distribution functions also depends on the factorization scheme.  The change of the hard scattering cross section is compensated for by the change of the distribution functions.
We study this situation using a Monte Carlo calculation. 
 
 In a previous work,\cite{rf:1} an algorithm that can reproduce the scaling violation for flavor singlet partons is constructed to the next-to-leading logarithmic (NLL) order of QCD.   That  algorithm consists of a model based on the evolution of momentum distributions.   In this model, a scaling violation of the parton distributions is generated by using only information from the splitting functions of the parton branching vertices and input distributions at a given energy, $Q_0$.  
  It has been found that, in the $\overline{\rm MS}$ scheme,\cite{rf:2}  the method reproduces  the scaling violation of the flavor singlet parton distributions, up to their normalizations.  

  In the model, the total momentum of the initial state partons is conserved.  Furthermore, it is not necessary to introduce non-trivial weight factors in order to reproduce the energy scale dependence of the momentum distribution of partons in the initial state. 
  These are advantages of this type of model in the calculation of the parton evolution to the NLL order of QCD.

 The  purpose of this paper is to extend the parton shower model proposed in Ref. \citen{rf:1} in order to generate parton evolution with any factorization scheme.
   In $\S$2, the algorithm for the mass singularity factorization of the parton shower model is explained.  As an example, the explicit expressions for the jet-calculus scheme are presented in $\S$3. 
 In $\S$4,  some numerical results obtained using the proposed method are presented.
Section 5 contains a summary.

\section{Collinear parton factorization}
 
 The parton evolution depends on the subtraction scheme used for the mass singularity, which appears due to collinear parton production.  
 In the $D=4-2\epsilon$ dimensional integration of the phase space, the factorized  contribution in $O(\alpha_s)$ with a factorization scheme $F$ can be written 
 \begin{eqnarray}
 {\hat F}^{[F]}_{ij}(\epsilon,z,M^2/\mu^2)\equiv {\hat F}^{[\overline{\rm MS}]}_{ij}(\epsilon,z)+{\hat P}_{ij}^{(0)}(z)\log\left({M^2 \over \mu^2}\right)+\Delta{\hat F}_{ij}^{[F]}(z),
   \end{eqnarray}
   with
     \begin{eqnarray}
{\hat F}^{[\overline{\rm MS}]}_{ij}(\epsilon,z)=\left({1 \over -\epsilon}-\log4\pi+\gamma_E\right){\hat P}_{ij}^{(0)}(z),
     \end{eqnarray}
where $i$ and $j$ indicate the types of partons ($q$ for a flavor singlet quark and $g$ for a gluon). Here, $M$ denotes the factorization scale of the collinear contributions.  In $4-2\epsilon$ dimensions, the strong coupling constant is defiend by $\alpha_s\mu^{2\epsilon}$, with the dimensionless coupling $\alpha_s$. The mass scale $\mu$ is also identified with the renormalization scale of the strong coupling constant.  Here, $z$ is the momentum fraction of the daughter parton and  $\gamma_E$ is the Euler constant. 
  Finally, ${\hat P}_{ij}^{(0)}(z)$ represents the infra-red unregulated splitting functions at the leading logarithmic (LL) order of QCD. \cite{rf:3} For example, $\Delta{\hat F}^{[\overline{\rm MS}]}_{ij}(z)=0$, with $\mu^2=M^2$, in Eq. (2$\cdot$1) corresponds to the factorized term with the $\overline{\rm MS}$ subtraction scheme.\cite{rf:2}
   
    In order to preserve the total momentum of the initial state partons, we subtract the infra-red regulated terms 
\begin{eqnarray}
 F_{qq}^{[F]}(\epsilon,z,M^2/\mu^2)&=&{1 \over z}\left[z{\hat F}_{qq}^{[F]}(\epsilon,z,M^2/\mu^2)\right]_+ \nonumber \\
 &-& \delta(1-z)\int^1_0dyy{\hat F}_{gq}^{[F]}(\epsilon,z,M^2/\mu^2), \\
 F_{gq}^{[F]}(\epsilon,z,M^2/\mu^2)&=&{\hat F}_{gq}^{[F]}(\epsilon,z,M^2/\mu^2),\\
 F_{qg}^{[F]}(\epsilon,z,M^2/\mu^2)&=&{\hat F}_{qg}^{[F]}(\epsilon,z,M^2/\mu^2), \\
 F_{gg}^{[F]}(\epsilon,z,M^2/\mu^2)&=&{1 \over z}\left[z{\hat F}_{gg}^{[F]}(\epsilon,z,M^2/\mu^2)\right]_+ \nonumber \\
 &-& \delta(1-z)\int^1_0dyy2N_f{\hat F}_{qg}^{[F]}(\epsilon,z,M^2/\mu^2) 
\end{eqnarray}
 from the hard scattering cross section. 
 Here $N_f$ denotes the number of flavors, and
\begin{eqnarray}
 \left[{\hat f}(z)\right]_+={\hat f}(z)-\delta(1-z)\int^1_0dy{\hat f}(y) 
\end{eqnarray}
for an unregulated function ${\hat f}(z)$ at $z=1$.
>From Eqs. (2.3)--(2.7), we find that
\begin{eqnarray}
 \int^1_0dzz\left[F_{qq}^{[F]}(\epsilon,z,M^2/\mu^2)+F_{gq}^{[F]}(\epsilon,z,M^2/\mu^2)\right] = 0
\end{eqnarray}
and 
\begin{eqnarray}
\int^1_0dzz\left[2N_fF_{qg}^{[F]}(\epsilon,z,M^2/\mu^2)+F_{gg}^{[F]}(\epsilon,z,M^2/\mu^2)\right]=0 
\end{eqnarray}
are satisfied for any factorization scheme $F$.
   
 Although $M$ and $\mu$ are independent parameters, we set $M=\mu$ for simplicity in the following calculation.  

   Including the NLL order terms, the splitting functions, $P_{ij}^{[F]}(\alpha_s,z)$, are given by 
\begin{eqnarray}
  P_{ij}^{[F]}(\alpha_s,z) = {\alpha_s(M^2) \over 2\pi}P^{(0)}_{ij}(z) 
    +  \left({\alpha_s(M^2) \over 2\pi}\right)^2 P^{[F](1)}_{ij}(z),  
  \end{eqnarray}
 where $ P^{(0)}_{ij}(z)$ and $ P^{[F](1)}_{ij}(z)$ are the infra-red regulated splitting functions of the LL order and of the NLL order, respectively, calculated with a factorization scheme $F$. 
Using
\begin{eqnarray}
\Delta F_{ij}^{[F]}(z)=F_{ij}^{[F]}(\epsilon,z,1)-F_{ij}^{[\overline{\rm MS}]}(\epsilon,z)
\end{eqnarray}
with $\epsilon \rightarrow 0$, the splitting functions to the NLL order are given by
\begin{eqnarray}
 P_{ij}^{[F](1)}(z)=P_{ij}^{[\overline{\rm MS}](1)}(z)-{\beta_0 \over 2}\Delta F_{ij}^{[F]}(z).
\end{eqnarray}
 Here, $\beta_0=11-2/3N_f$, and $P_{ij}^{[\overline{\rm MS}](1)}(z)$ is the splitting function calculated with the $\overline{\rm MS}$ scheme.\cite{rf:4}

The quantities 
\begin{eqnarray}
  P_{q}^{[F]} (\alpha_s,z)&=&P_{qq}^{[F]} (\alpha_s,z) +  P_{gq}^{[F]} (\alpha_s,z)
  \end{eqnarray}
and
 \begin{eqnarray}
  P_{g}^{[F]} (\alpha_s,z)&=&2N_fP_{qg}^{[F]} (\alpha_s,z) + P_{gg}^{[F]} (\alpha_s,z)
    \end{eqnarray}
 satisfy 
\begin{eqnarray}
   \int^{1}_{0} dz zP_{i}^{[F]}(\alpha_s,z) = 0
\end{eqnarray}
for $i=q,g$. This guarantees that the total momentum of the initial state hadron is conserved, that is,
\begin{eqnarray}
    \int^1_0 dx x \left[ \Sigma_F(x,K^2) + G_F(x, K^2) \right] = 1
\end{eqnarray}
for any $ K^2 $, where $x$ is the momentum fraction of a parton inside the hadron.  Here, $ \Sigma_F(x,K^2) $ and $ G_F(x,K^2) $ are the particle number distribution function of the flavor singlet quarks and that of the gluons in the factorization scheme $F$, respectively. 

Using 
\begin{eqnarray}
\int^1_{1-\delta}dzzP_i^{[F]}(z)=-\int^{1-\delta}_0dzz{\hat P}_i^{[F]}(z),
\end{eqnarray}
 the non-branching probablities for the flavor singlet quarks $(i=q)$ and for the gluons $(i=g)$ are defined by
\begin{eqnarray}
\Pi_{NB}^{[F](i)} ( K^2_2, K^2_1 ) = {\rm exp} \left[ - \int^{K^2_2 }_{K^2_1} {dK^2 \over K^2 } \int^{1 - \delta}_0 dzz{\hat P}_{i}^{[F]} (\alpha_s,z)  \right]. 
\end{eqnarray}
Here, the partons inside an initial state hadron have space-like virtualities (i.e., $k_i^2 \equiv -K_i^2 <0$), and $\delta$ denotes the resolution of the momentum fraction of the final state partons.  
 Although in the Monte Carlo calculation, only unregulated splitting functions are used, the infra-red regulated terms presented in Eqs. (2.3)--(2.6) must be subtracted from the hard scattering cross section in order to cancel the dependence on the factorization scheme in the hadronic cross section.
   
 Using this algorithm, the momentum distributions are reproduced up to their normalizations without the introduction of any non-trivial weight factor. 
Therefore, the model guarantees the total normalization of the momentum distributions for the flavor singlet partons in any factorization scheme.

\section{ Factorization in the jet calculus scheme }

  In the jet calculus (JC)\cite{rf:5}, the singular terms due to the collinear parton radiation, which are integrated over the $t$-channel momentum of the radiated parton, with the range of integration $0 \leq -{\hat t} \leq M^2$, are given by
  \begin{eqnarray}
 {\hat F}^{[{\rm JC}]}_{ij}(\epsilon,z,M^2/\mu^2)={{\hat P}_{ij}(z,\epsilon)\over \Gamma(1-\epsilon)}\left[{4\pi\mu^2 \over M^2}\right]^{\epsilon}{(1-z)^{-\epsilon} \over -\epsilon}, 
 \end{eqnarray}
 with
 \begin{eqnarray}
  {\hat P}_{ij}(z,\epsilon)={\hat P}_{ij}^{(0)}(z)+\epsilon {\hat P}'_{ij}(z)  \end{eqnarray}
  for $i,j=q,g$.
 Here, the functions ${\hat P}'_{ij}(z)$ are given in Ref. \citen{rf:6}.
 
 From Eq. (2$\cdot$11), the functions $\Delta F_{ij}^{[{\rm JC}]}(z)$ in the JC scheme are given by
\begin{eqnarray}
  \Delta F_{qq}^{[{\rm JC}]}(z)&=&\left[{\hat P}^{(0)}_{qq}(z)\log(1-z)-{\hat P}'_{qq}(z)\right]_+-\delta(1-z){1 \over 2}C_F, \\
 \Delta F_{gq}^{[{\rm JC}]}(z)&=&{\hat P}^{(0)}_{gq}(z)\log(1-z)-{\hat P}'_{gq}(z),\\
 \Delta F_{qg}^{[{\rm JC}]}(z)&=&{\hat P}^{(0)}_{qg}(z)\log(1-z)-{\hat P}'_{qg}(z),\\
 \Delta F_{gg}^{[{\rm JC}]}(z)&=&{1 \over z}\left[z{\hat P}^{(0)}_{gg}(z)\log(1-z) \right]_++\delta(1-z)2N_fT_R{29 \over 72}, 
\end{eqnarray}
with
$C_F=4/3$ and $T_R=1/2$.

In the Monte Carlo calculation, we study the following two cases.
\vspace{3mm}
\begin{description}

\item{(Case 1)}: The splitting functions 
 \begin{eqnarray}
  {\hat P}_{ij}^{[{\rm JC}1]}(\alpha_s,z) = {\alpha_s(M^2) \over 2\pi}{\hat P}^{(0)}_{ij}(z) 
    +  \left({\alpha_s(M^2) \over 2\pi}\right)^2{\hat P}^{[{\rm JC}1](1)}_{ij}(z),  
  \end{eqnarray}
  with
  \begin{eqnarray}
 {\hat P}_{ij}^{[{\rm JC}1](1)}(z)={\hat P}_{ij}^{[\overline{\rm MS}](1)}(z)-{\beta_0 \over 2}\Delta{\hat F}_{ij}^{[{\rm JC}]}(z),
\end{eqnarray}
are used in the non-branching probabilities, and the momentum fraction $z$ is generated according to the probability $z{\hat P}_{ij}^{[{\rm JC}1]}(\alpha_s,z)$. 

\vspace{3mm}

\item{(Case 2)}: $\Delta{\hat F}_{qq}^{[{\rm JC}]}$ and $\Delta{\hat F}_{gg}^{[{\rm JC}]}$ behave as $ \sim \log(1-z)/(1-z) $ for $z \rightarrow 1$. This may yield large higher-order contributions for $z \sim 1$. 
Such dangerous terms can be absorbed into the strong coupling constant\cite{rf:5} as 
\begin{eqnarray}
 \alpha_s(K^2) \rightarrow \alpha_s\left(f_{ij}(z)(1-z)K^2\right),
\end{eqnarray}
with
\begin{eqnarray}
f_{ij}(z)=\exp\left[{{\hat P}'_{ij}(z) \over {\hat P}_{ij}^{(0)}(z) }\right].
\end{eqnarray}
The splitting functions in the JC scheme are modified as
\begin{eqnarray}
{\hat P}_{ij}^{[{\rm JC}2]}(\alpha_s,z) = {\alpha_s\left({\bar K}^2_{ij}\right) \over 2\pi}{\hat P}^{(0)}_{ij}(z)   +  \left({\alpha_s\left({\bar K}^2_{ij}\right) \over 2\pi}\right)^2{\hat P}^{\overline{\rm MS}(1)}_{ij}(z), 
\end{eqnarray}
with ${\bar K}^2_{ij}=f_{ij}(z)(1-z)K^2 \simeq (0.4 \sim 1)p_T^2$, where $p_T$ is the transverse momentum of the generated parton with space-like virtuality.  
The factors $f_{ij}(z)$ are presented in Fig. 1.
\begin{figure}
\centerline{\includegraphics[width=10cm]{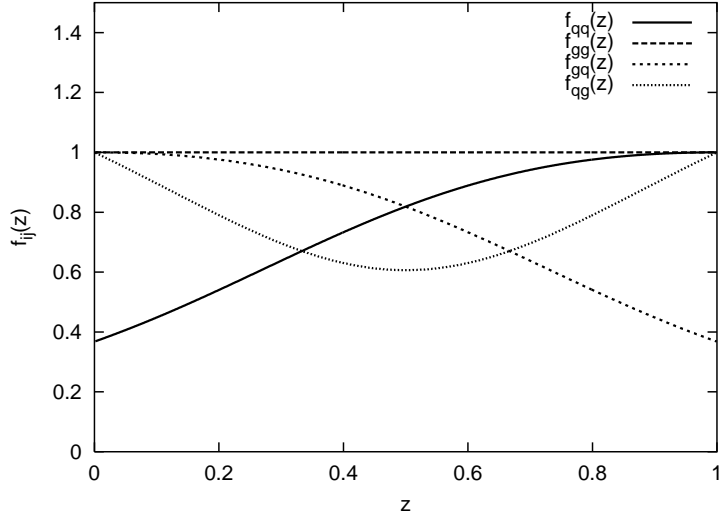}}
\caption{The factors $f_{ij}(z)$ defined in Eq. (3.10).}
\end{figure}

\end{description}

 \section{ Numerical results for the JC scheme }
 
 In this section, some numerical results obtained using the algorithm explained in the previous sections are presented. 
 
 In order to check the scaling violation generated by the Monte Carlo calculation with the JC scheme, we calculate the momentum distribution functions in an $O(\alpha_s)$ approximation.
 The momentum distribution functions in the JC scheme are given by 
 \begin{eqnarray}
 x\Sigma_{\rm JC}(x,M^2) &=& x\Sigma_{\overline{\rm MS}} (x,M^2) +{\alpha_s(M^2) \over 2\pi}\int^1_x{dz \over z}z\left[\Delta F_{qq}^{[{\rm JC}]}(z){x \over z}\Sigma_{\overline{\rm MS}}\left({x \over z},M^2\right) \right.\nonumber \\ 
  &+& \left.\Delta F_{qg}^{[{\rm JC}]}(z){x \over z}G_{\overline{\rm MS}} \left({x \over z},M^2\right) \right] 
   \end{eqnarray}
and 
 \begin{eqnarray}
 xG_{\rm JC}(x,M^2) &=& xG_{\overline{\rm MS}} (x,M^2) +{\alpha_s(M^2) \over 2\pi}\int^1_x{dz \over z}z\left[\Delta F_{gq}^{[{\rm JC}]}(z){x \over z}\Sigma_{\overline{\rm MS}}\left({x \over z},M^2\right) \right.\nonumber \\
&+& \left. \Delta F_{gg}^{[{\rm JC}]}(z){x \over z}G_{\overline{\rm MS}} \left({x \over z},M^2\right) \right]. 
\end{eqnarray}
  The distribution functions $x\Sigma_{\overline{\rm MS}} (x,M^2)$ and $xG_{\overline{\rm MS}}(x,M^2)$ are obtained with GRV(98)\cite{rf:7}.

 In the Monte Carlo calculation, the parton showers  start from the momentum distributions at $M_0^2=5~{\rm GeV}^2$. In Fig. 2, the momentum distribution functions for the gluon at $M^2=10^3~{\rm GeV}^2$  are indicated  by  `$\Box$' for case 1 and by `$+$'  for case 2 in the JC scheme, with  $ \delta = l_0^2/K^2 (l_0^2=0.2~{\rm GeV}^2)$.

  Here, the results obtained from the Monte Carlo calculation are plotted as 
$dN/dy_F$, with $ y_F = -{\rm ln}x_F $, where $dN$ denotes the average number of events generated within the range $y_F \pm dy_F/2$.  The quantity $dN$ divided by $dy_F$ corresponds to the momentum distribution multiplied by $ x_F $ [i.e., $x_F(x_Ff(x_F,M^2))$].

  In the figure, the dotted curve and the dashed curve are  the results obtained with GRV(98) in the $\overline{\rm MS}$ scheme for $M^2_0=5~{\rm GeV}^2$ and $M^2=10^3~{\rm GeV}^2$, respectively.  The dash-dotted curve and the solid curve denote the distribution functions calculated using Eq. (4.2) for $M^2_0=5~{\rm GeV}^2$ and $M^2=10^3~{\rm GeV}^2$, respectively. 
  In Fig. 3, similar results for the flavor singlet quarks are also presented.

  \begin{figure}
\centerline{\includegraphics[width=10cm]{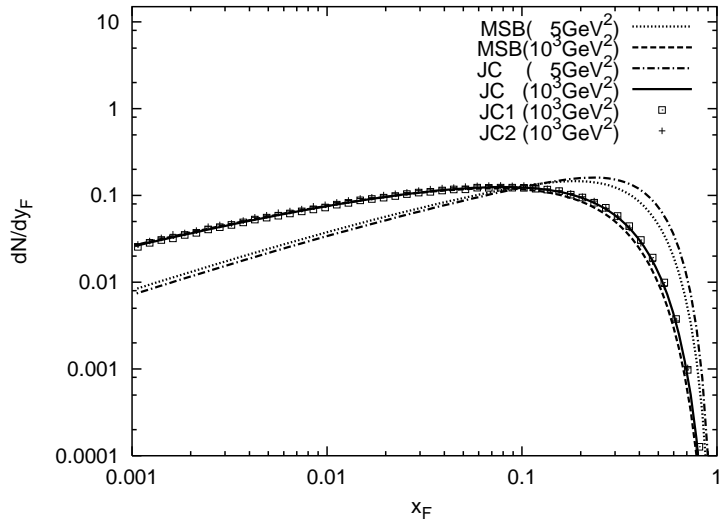}}
\caption{The momentum distribution functions (multiplied by $x_F$) for the gluon  obtained with the algorithm explained in $\S$$\S$2 and 3.  
The results obtained from the Monte Carlo simulation are indicated by `$\Box$' for case 1 and `$+$' for case 2 in the JC scheme with $l_0^2=0.2~{\rm GeV}^2$ at $M^2=10^3~{\rm GeV}^2$.  
 The solid curve and  dashed curve represent the corresponding results for the JC scheme and for the $\overline{\rm MS}$ scheme obtained from GRV(98).  The input distributions at $M_0^2=5~{\rm GeV}^2$ are represented by the dotted curve for the $\overline{\rm MS}$ scheme and the dash-dotted curve for the JC scheme.  }
\end{figure}
\begin{figure}
\centerline{\includegraphics[width=10cm]{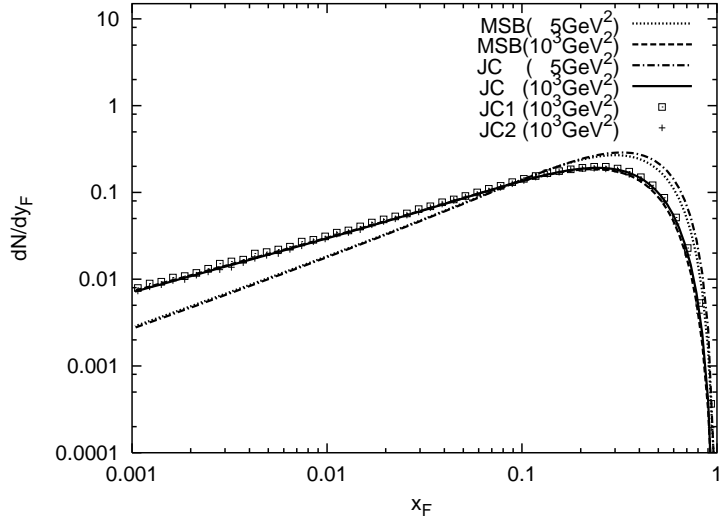}}
\caption{The momentum distribution functions (multiplied by $x_F$) for the flavor singlet quarks with the algorithm explained in $\S$$\S$2 and 3.  The notation is the same as in Fig. 2.}
\end{figure}

\begin{figure}
\centerline{\includegraphics[width=10cm]{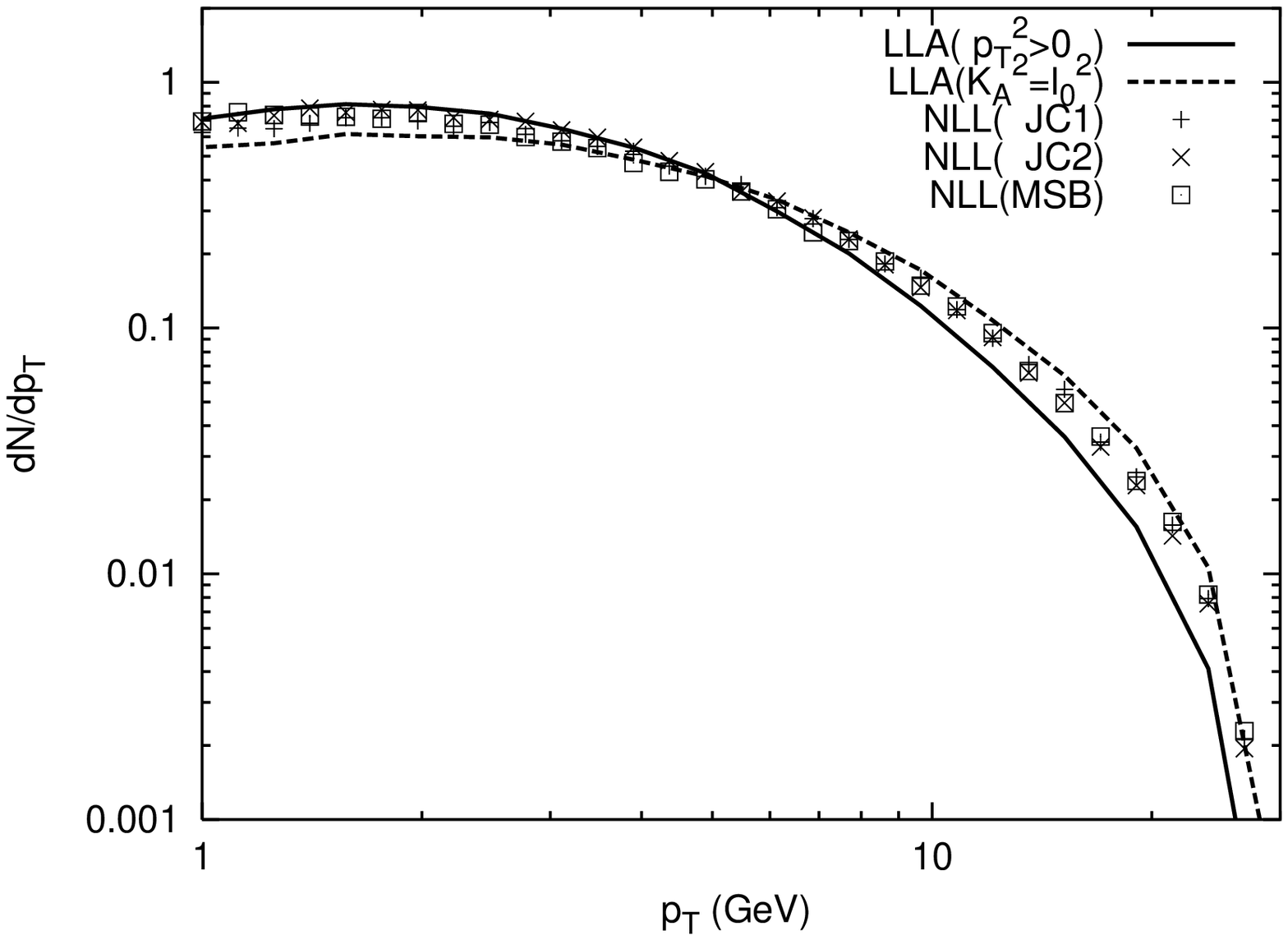}}
\caption{The transverse momentum distributions for the gluons with final momentum fractions satisfying $0.09 \leq x_F \leq 0.11$ at $M^2=10^3~{\rm GeV}^2$.  The results obtained from the Monte Carlo simulation are represented by `$+$' for case 1, `$\times$' for case 2 in the JC scheme with $\l_0^2 \leq K_A^2 \leq f_A(-s)$, and `$\Box$' for the $\overline{\rm MS}$ scheme. The solid curve and the dashed curve represent the phase space condition due to the relation $p_T^2 \geq 0$ and the radiation of partons with fixed $K_A^2=l_0^2~(=0.5~{\rm GeV}^2)$ in the LL  order of QCD, respectively.}
\centerline{\includegraphics[width=10cm]{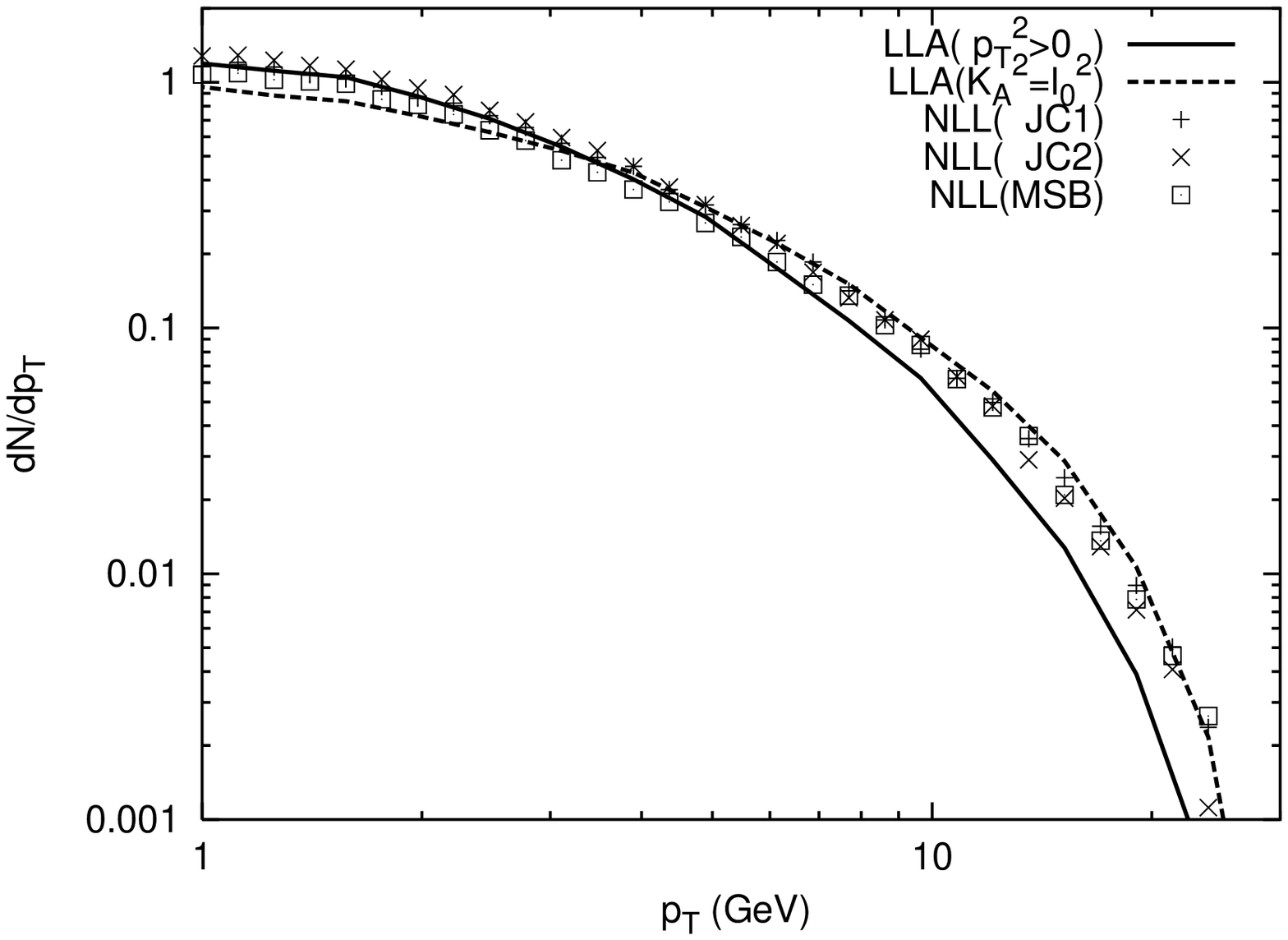}}
\caption{The transverse momentum distributions for the flavor singlet quarks with final momentum fractions satisfying $0.09 \leq x_F \leq 0.11$ at $M^2=10^3~{\rm GeV}^2$.  The notation is the same as in Fig. 4. }
\end{figure}

Next, we present the transverse momentum distributions of the generated partons in the JC scheme.  
The transverse momenta for the partons are restricted by the NLL order terms.\cite{rf:1}  At the LL order of QCD, the upper limit of the virtuality for the generated parton is restricted only by a large energy scale, which contributes to the branching vertex. In the actual Monte Carlo generation, the allowed kinematic boundary for the two-body branching is determined by the condition that the squared transverse momenta of the generated partons be positive.  

In Ref. \citen{rf:1}, the transverse momenta of the initial state partons with space-like virtuality are generated using the effective two body vertices, where the three-body decay functions calculated in Refs. \citen{rf:8} and \citen{rf:9}\footnote{The function $F(x)$ in the appendix of Ref. \citen{rf:9} is given by $F(x)=x^2+(1-x)^2$.} are included as the boundary conditions of the virtuality for the out-going partons.  

  The transverse momentum of the parton with momentum $k_3$ for the branching $a(p) \rightarrow b(k_1+k_2)+c(k_3)$ is given by
 \begin{eqnarray}
  p_T^2=x_3y_3\left[ p^2 + {-s \over x_3} - {K_A^2 \over y_3}\right], 
   \end{eqnarray} 
where $K_A^2=(k_1+k_2)^2 (\geq 0), s=k_3^2(<0)$ and $p_T^2={\vec k_{3T}}^2$, with $y_3=1-x_3$. Here, $x_3$ is the momentum fraction of the parton with momentum $k_3$. The virtuality $K_A^2$ is restricted by the condition $K_A^2 \leq f_A(-s)$ at the NLL order, instead of $K_A^2 \leq y_3/x_3(-s)$, by the kinematic boundary of the two-body branching, due to the fact that $p^2_T \geq 0$ for $-p^2 \ll -s, K_A^2$. Here, the factor of $f_A$ for the two-gluon radiation processes is given in Ref. \citen{rf:1}. 

In Fig. 4, the transverse momentum distributions for the gluons with final momentum fractions satisfying $0.09 \leq x_F \leq 0.11$ are presented for $M^2=10^3~{\rm GeV}^2$.  The results obtained from the Monte Carlo simulation are represented by `$+$' for case 1, `$\times$' for case 2 in the JC scheme, and `$\Box$' for the $\overline{\rm MS}$ scheme, with $\l_0^2 \leq K_A^2 \leq f_A(-s)$. The solid curve and the dashed curve correspond to the case $\l_0^2 \leq K_A^2 \leq y_3/x_3(-s)$ due to $p_T^2 \geq 0$ and the parton radiations with fixed $K_A^2=l_0^2~(=0.5~{\rm GeV}^2)$ at the LL order, respectively.  In Fig. 5, similar results for the flavor singlet quarks are presented.\footnote{The transverse momentum distributions with $l^2_0=1~{\rm GeV}^2$ are presented in Figs. 6 and 7 of Ref. \citen{rf:1}.  }
 
\section{Summary}
 
   In this paper, we studied a factorization algorithm for the parton shower model based on the evolution of the momentum distributions, which was proposed in a previous paper.  In this algorithm, we subtracted the singular terms due to the collinear parton production from the hard scattering cross section so as to
 preserve  the total momentum of the initial state partons with any choice of the factorization scheme.

The results obtained with the jet calculus scheme were compared with those obtained using the corresponding distribution functions calculated in an $O(\alpha_s)$ approximation.  These two methods give consistent results for the scaling violation of the flavor singlet parton distributions.  The algorithm developed in this paper reproduces the dependence on the factorization scheme of the parton distribution functions.  

For the transverse momentum distributions, in the large $p_T$ region, we find that the dependence on the factorization scheme at the NLL order is smaller than the ambiguity resulting from the choice of the kinematic boundary at the LL order of QCD.

The method presented in this paper may be useful in the construction of Monte Carlo generators beyond the LL order of QCD, which are desired for future high energy experiments, such as LHC at CERN.

\section*{Acknowledgements}

This work was supported in part by RCMAS (the Research Center for Measurement in Advanced Science) of Rikkyo University and the Japan Society for the Promotion of Science through a Grant-in-Aid for Scientific Research B (No. 13135220).


\end{document}